# Data integrity vs. inference accuracy in large AIS datasets

Adam Kiersztyn[1][0000−0001−5222−8101], Dariusz Czerwinski[1][0000−0002−3642−1929], Aneta Oniszczuk-Jastrzabek[2][0000−0002−4268−0011], Ernest Czermanski[2][0000−0002−2114−8093], and Agnieszka Rzepka[3][0000−0003−4495−6066]

[1] Lublin University of Technology, Faculty of Mathematics and Information Technology, Lublin, Poland
`a.kiersztyn@pollub.pl`
[2] University of Gdansk, Department of Maritime Transport and Seaborne Trade, Gdansk, Poland
[3] Lublin University of Technology, Faculty of Management, Lublin, Poland

**Abstract.** Automatic Ship Identification Systems (AIS) play a key role in monitoring maritime traffic, providing the data necessary for analysis and decision-making. The integrity of this data is fundamental to the correctness of inference and decision-making in the context of maritime safety, traffic management and environmental protection. This paper analyzes the impact of data integrity in large AIS datasets, on classification accuracy. It also presents error detection and correction methods and data verification techniques that can improve the reliability of AIS systems. The results show that improving the integrity of AIS data significantly improves the quality of inference, which has a direct impact on operational efficiency and safety at sea.

## 1 Introduction

As part of data integrity testing, a very important issue is the detection of outlier data and anomalies. Among the many methods for detecting outliers, it is worth mentioning Isolation Forest [1] and its numerous modifications [2] [3]. In addition, it is worth mentioning other methods that allow for detecting outliers in multimodal data sets [4]. An interesting approach is the detection of anomalies based on information granules [5] [6].
In coastal regions, the littoral AIS falls short of ensuring operational continuity and system availability, leaving certain areas uncovered by the network. The authors in the paper [7] propose methods to monitor the integrity of AIS dynamic data using process models like GPS, Dead Reckoning, and RADAR EKF-SLAM. The reliability of AIS data was assessed using stochastic techniques grounded in Markov chains.

The paper [8] presents algorithms to improve the quality and integrity of AIS data for ship trajectories. The authors presented algorithms for error pre-processing, focusing on physical integrity, spatial logical integrity, and time accuracy. To verify applicability, track comparison maps and traffic density maps for various ship types were generated using AIS data from the Chinese Zhoushan Islands.

Iphar et al. in [9] developed methodologies to assess AIS message integrity and veracity, employing rule-based methods and logic-based frameworks to detect anomalies and trigger situation-specific alerts.

A new method for identifying and repairing abnormal points in trajectories only based the AIS data of the ship itself, which can effectively reduce the missed judgment of outliers is proposed in [10].

The paper [12] proposes a framework to reconstruct accurate ship trajectories from noisy AIS data using data quality control and prediction. The framework's data quality control involved three steps: separating trajectories, de-noising data, and normalizing it. Outliers in raw AIS data were removed using a moving average model, then normalized.

The lack of data integrity leaves AIS messages open to unauthorized alterations as shown in [13]. Interestingly, the concept of non-repudiation for AIS messages has only recently garnered attention, despite its crucial role in probing maritime incidents and breaches of maritime regulations.

The authors in [14] pointed that straightforward nature of the AIS protocol has enabled its use in numerous applications today. However, AIS still lacks the ability to verify message integrity and authentication. For navigators, it is crucial to cross-check AIS information with radar and visual observations. A significant safety concern in navigation is AtoN spoofing, particularly with V-AtoNs.

Since AIS was introduced to maritime transportation, its potential practical uses have been increasingly acknowledged in academic circles as stated in [15]. Initially, this recognition focused on AIS data mining and its applications for navigation safety. As data quality and accessibility improved, research expanded beyond navigation safety to explore broader and more advanced uses of AIS data.

Previous studies on ship traffic using AIS data have demonstrated that analyzing historical data can yield valuable insights into ship behavior. Additionally, the probabilistic characterization of ship trajectories along specific routes facilitates real-time anomaly detection, which is crucial for developing alerts to aid in traffic supervision and control. Quality of AIS data in such cases is crucial as pointed in [16].

Identifying unusual vessel behavior, basing on the AIS data, to advance autonomous vessel technology for sustainable marine transportation was presented in [17]. Authors in [18] stated that detecting anomalies, such as unexpected sailing behavior, in vessel trajectories is critically important. Methods for identifying these anomalies range from developing normality models to pinpointing specific incidents, like AIS switch-offs or collision avoidance maneuvers.

In the article [19] authors introduce a rule-based method for assessing data integrity. The rules are derived from system technical specifications and expert knowledge, and are formalized using a logic-based framework, which triggers situation-specific alerts. Detecting abnormal data in AIS systems and databases aims to enhance marine traffic surveillance, safeguard human life at sea, and mitigate hazardous behaviors affecting ports, offshore structures, and the environment.

The study of Wang et al. in [20] introduces a vision-inspired framework to classify AIS data quality issues, addressing the limitations of traditional statisti- cal methods in optimizing maritime operations. Accurate data quality diagnosis is essential for trustworthy AI-driven decision-making.

The paper [21] presents the results of a spatial analysis of a large volume of AIS data aimed at detecting predefined maritime anomalies. These anomalies are categorized into traffic analysis, static anomalies, and loitering detection. The analysis utilized advanced algorithms and technology capable of efficiently processing big data.

Efficiently managing AIS data is crucial for improving maritime safety and navigation, but it is challenged by the system's large volume and error-prone datasets. Authors in paper [22] introduces the Automatic Identification System Database (AISdb), a new tool developed to tackle the difficulties in processing and analyzing AIS data.

This article [23] examines the landscape of global navigation satellite system (GNSS) spoofing. While it is well known that automated identification system (AIS) spoofing can be used in electronic warfare to hide military activities in sensitive sea areas, recent events indicate a growing interest in spoofing AIS signals for commercial purposes. The shipping industry is currently facing an unprecedented wave of deceptive practices by tanker operators attempting to evade sanctions. These false ship positions highlight the urgent need for effective tools and strategies to ensure the reliability and robustness of AIS.

In the works [24] and [25] authors pointed that quality of large AIS datasets is very important for proper analysis. AIS data is crucial for enhancing the safety, efficiency, environmental performance, and operations of the global shipping in- dustry. This investigation examines four key aspects of data quality: accuracy, completeness, consistency, and timeliness. The findings reveal that the quality of marine AIS data is influenced by AIS technology, communication protocols, environmental conditions, and human factors.

These studies collectively aim to enhance maritime domain awareness, im- prove data quality, and mitigate risks associated with AIS data integrity issues. AIS data can lack integrity due to errors or intentional falsification, requiring a methodology to assess its veracity. In this study we proposed analysis of AIS data quality based on the database which was purchased on a commercial basis from the S&P Global data provider, within the IHS Markit. The database consists of more than 225 indicators describing a ship, where data is (or, at should be at least) sourced from the vessels registration authority. Chapter 2 consists the analysis of the data integrity for the purpose of inference accuracy. The article is finalized with conclusions.

## 2      Data analysis for integrity

The analysis of AIS data integrity was limited to one type of vessel, namely tankers. This type of vessel is characterized by relatively high variability and, unlike passenger ships, is not limited to fixed routes. Additionally, these vessels maintain relatively stable speeds. The recorded speed was analyzed in detail, as this variable most clearly demonstrates the presence of both outliers and more complex anomalies. A summary of all records pertaining to tankers by year is presented first (compare figure 1).

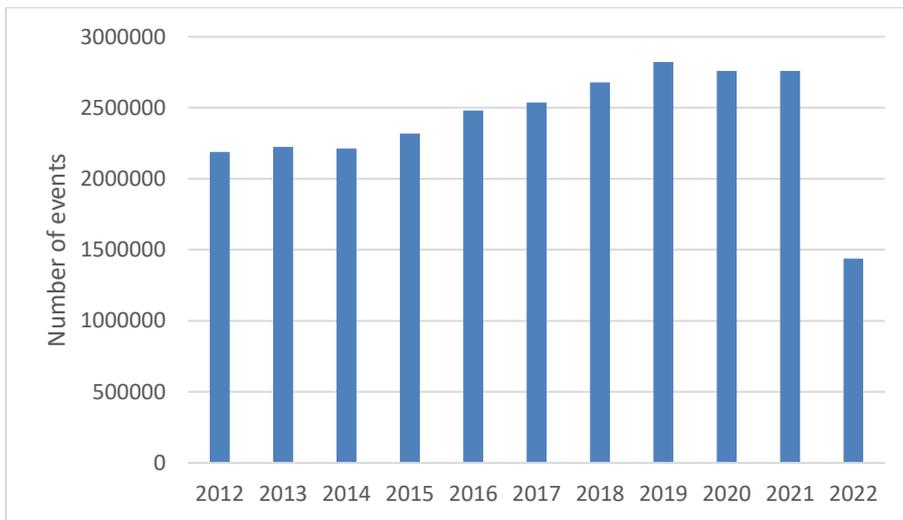

**Fig.1.** Number of records in the database describing ship traffic

In the last analyzed year, there is a significant drop in data. This is due to the fact that only half-year data for 2022 was analyzed. Analyzing the data presented in Figure 1, it can be observed that the number of events increased until the outbreak of the pandemic. After a brief decline, a clear rebound was noted.

It is worth comparing the number of records with the number for which movement was recorded. Figure 2 presents the number of different tankers in individual years. Comparing the number of observations with the number of ships, a significant correlation can be seen. This might suggest that a similar number of observations is recorded for each ship. However, this is not the case, as evidenced by Figure 3.

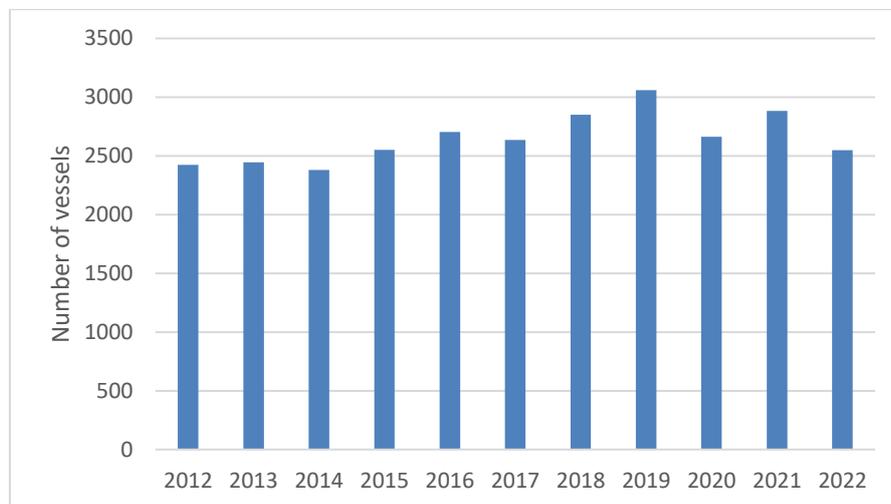

**Fig. 2.** Number of tankers sailing in the Baltic Sea by year

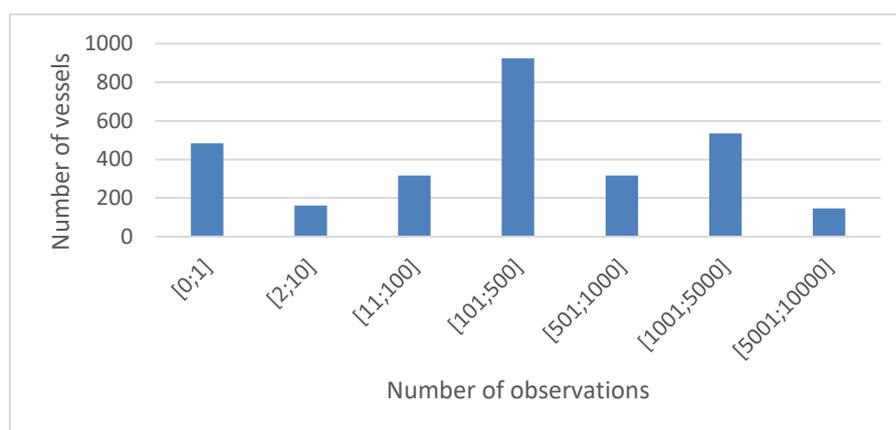

**Fig. 3.** Summary of the number of records for each ship

For a large number of tankers, only single observations are recorded, which may be due to the fact that the boundary between the Baltic Sea and the North Sea is arbitrary, and some tankers only briefly touch the Baltic. Moving on to more detailed analyses, it is worth comparing the recorded movement status for tankers in individual years. Such a comparison is presented in Table 1.

**Table 1.** Summary of "traffic statuses" by year

| Status\Year | 2012 | 2013 | 2014 | 2015 | 2016 | 2017 | 2018 | 2019 | 2020 | 2021 | 2022 |
|---|---|---|---|---|---|---|---|---|---|---|---|
| Aground | 233 | 591 | 71 | 226 | 38 | 372 | 290 | 166 | 158 | 78 | 553 |

| | | | | | | | | | | |
|---|---|---|---|---|---|---|---|---|---|---|
| Anchored | 342448 | 372491 | 369178 | 346248 | 424284 | 385736 | 434652 | 476774 | 528991 | 474696 | 256514 |
| Constrained by draught | 31051 | 29496 | 26767 | 27487 | 30252 | 27613 | 26564 | 26756 | 22553 | 22663 | 16847 |
| Engaged in fishing | 55 | 224 | 9 | 72 | 14 | 171 | 707 | 18 | 16 | 56 | 39 |
| Moored | 556459 | 539556 | 547706 | 572592 | 607352 | 663804 | 724851 | 769887 | 755480 | 783201 | 394528 |
| N/A | 65281 | 76899 | 75193 | 118676 | 160342 | 182812 | 193448 | 211266 | 210582 | 198546 | 87523 |
| Not under command | 12183 | 16766 | 11473 | 3769 | 2813 | 3247 | 3509 | 2548 | 1395 | 4202 | 2264 |
| Restriced manoeuvera-bility | 3565 | 1624 | 1130 | 1691 | 7144 | 3865 | 1857 | 1705 | 1225 | 1077 | 2268 |
| Under way sailing | 17651 | 6083 | 16055 | 27351 | 24234 | 33646 | 18764 | 19240 | 11896 | 11841 | 7355 |
| Under way using engine | 1157339 | 1178957 | 1163376 | 1218038 | 1221510 | 1233459 | 1271917 | 1311935 | 1225913 | 1261408 | 668563 |

Analyzing the data presented in Table 1, it is easy to notice certain anomalies and numerous "N/A" data gaps. Anomalies can certainly include "Not under command" or "Under way sailing," which seem logically unjustified for tankers. By limiting ourselves to two selected statuses, namely "Moored" and "Under way using engine," we can identify further anomalies and evident outliers.

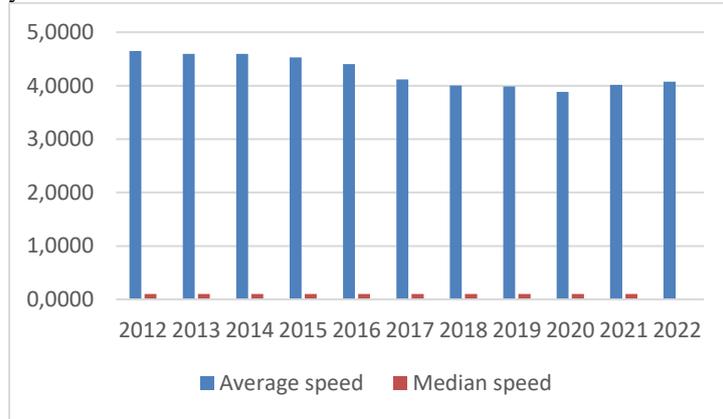

**Fig. 4.** Mean and median speeds for all available data

Such evident data integrity issues clearly confirm that proper preprocessing is essential before starting analyses. Otherwise, one might obtain models that could lead to erroneous conclusions.

**Table 2.** Summary of average and median speeds by year for two selected traffic statuses

| Year\Statistics | Moored Average | Moored Median | Engine Average | Engine Median |
|---|---|---|---|---|
| 2012 | 0,1748 | 0,0000 | 8,2631 | 10,7000 |
| 2013 | 0,1643 | 0,0000 | 8,1699 | 10,6000 |
| 2014 | 0,1714 | 0,0000 | 8,1116 | 10,5000 |
| 2015 | 0,1459 | 0,0000 | 7,9135 | 10,2000 |
| 2016 | 0,1280 | 0,0000 | 7,8792 | 10,1000 |
| 2017 | 0,1118 | 0,0000 | 8,0629 | 10,3000 |
| 2018 | 0,0970 | 0,0000 | 8,0141 | 10,3000 |
| 2019 | 0,0922 | 0,0000 | 8,1805 | 10,4000 |
| 2020 | 0,0885 | 0,0000 | 8,3534 | 10,5000 |
| 2021 | 0,0736 | 0,0000 | 8,4195 | 10,6000 |
| 2022 | 0,0680 | 0,0000 | 8,3429 | 10,5000 |

Analyzing the mean and median speeds (see Figure 4), it is easy to observe that the means are significantly higher than the medians. This is closely related to the fact that the mean is very sensitive to outliers. Therefore, in the case of average speed, one should expect many values significantly higher than the others, which may suggest that these are outliers. Indeed, in the case of tanker speeds, the maximum value is as high as 102 knots.

An interesting point in Table 2 is the average speed values in the 'Moored' status. Positive values suggest that a significant portion of observations have a speed value greater than zero, which is completely physically unjustified for the 'Moored' status. Therefore, we are dealing with evident anomalies. In the case of the second status, the median value exceeds the average value each year, suggesting that there are numerous observations with values close to zero.

The analysis of the data presented in Figure 5 allows for the observation of several significant anomalies. Firstly, many ships switch to the 'Moored' status in open sea, which is physically impossible. Additionally, moored ships cannot move at a speed of 102 knots towards the shore. Besides anomalies and outliers in the AIS data, data gaps can also be encountered. The scale of this problem is illustrated by the example ship route shown in Figure 6.

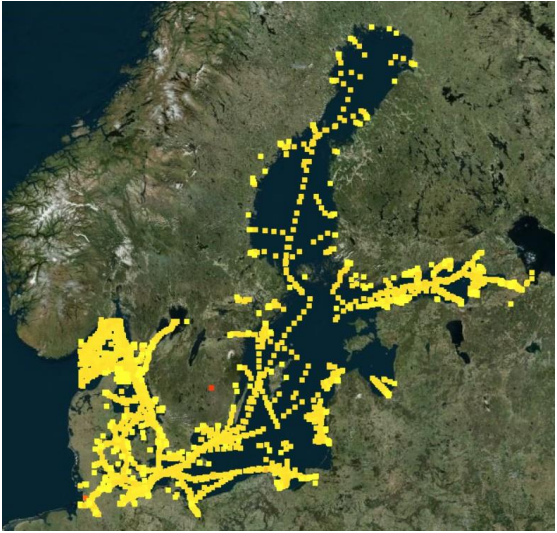

**Fig. 5.** Positions of ships of "Moored" status in 2017 with marked speeds. Maximum values (102 knots) are marked in red.

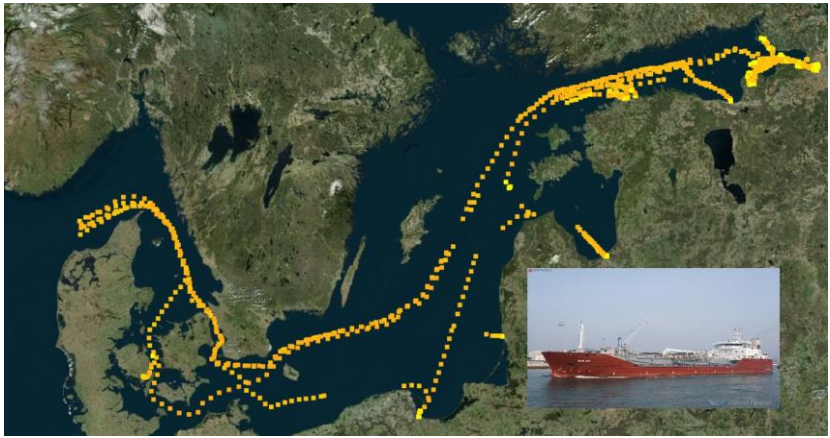

**Fig. 6.** Positions of the ship named GOGLAND in the years 2012 -2022

## 3  Conclusions

The study showed that AIS data integrity is crucial for accurate inferences and safe management of maritime traffic. Problems related to incompleteness and anomalies in AIS data can lead to incorrect operational decisions, which threat- ens safety at sea. The analysis showed that error detection and correction tech- niques and data verification methods can significantly improve the quality of AIS data, thus reducing the risk associated with their incorrect interpretation. The results confirm that proper data processing before analysis is essential to avoid false conclusions, especially in the context of the development of autonomous maritime technologies and advanced maritime traffic monitoring systems.

## 4  Acknowledgements

The work was co-financed by the Lublin University of Technology Scientific Fund: FD-20/IT-3/002

13. Goudossis, A. & Katsikas, S. Towards a secure automatic identification system (AIS). *Journal Of Marine Science And Technology (Japan)*. **24**, 410-423 (2019)
14. Androjna, A., Perkovič, M., Pavic, I. & Mišković, J. Ais data vulnerability indi- cated by a spoofing case-study. *Applied Sciences (Switzerland)*. **11**, 5015 (2021)
15. Yang, D., Wu, L., Wang, S., Jia, H. & Li, K. How big data enriches maritime re- search–a critical review of Automatic Identification System (AIS) data applications. *Transport Reviews*. **39**, 755-773 (2019)
16. Rong, H., Teixeira, A. & Guedes Soares, C. Data mining approach to shipping route characterization and anomaly detection based on AIS data. *Ocean Engineering*. **198** pp. 106936 (2020)
17. Han, X., Armenakis, C. & Jadidi, M. Modeling vessel behaviours by clustering ais data using optimized dbscan. *Sustainability (Switzerland)*. **13**, 8162 (2021)
18. Kontopoulos, I., Varlamis, I. & Tserpes, K. A distributed framework for extract- ing maritime traffic patterns. *International Journal Of Geographical Information Science*. **35**, 1-26 (2020)
19. Iphar, C., Ray, C. & Napoli, A. Data integrity assessment for maritime anomaly detection. *Expert Systems With Applications*. **147** pp. 113219 (2020)
20. Wang, K., Tristan, O., Zhang, X., Fu, X. & Qin, Z. Data-centric AI practice in maritime: Securing trusted data quality via a computer vision-based framework. *Proceedings - 2024 IEEE Conference On Artificial Intelligence, CAI 2024*. pp. 414- 417 (2024)
21. Filipiak, D., Strozyna, M., Wecel, K. & Abramowicz, W. Big Data for Anomaly De- tection in Maritime Surveillance: Spatial AIS Data Analysis for Tankers. *Maritime Technical Journal*. **215**, 5-28 (2019)
22. Spadon, G., Kumar, J., Chen, J., Smith, M., Hilliard, C., Vela, S., Gehrmann, R., DiBacco, C., Matwin, S. & Pelot, R. Maritime Tracking Data Anal- ysis and Integration with AISdb. *ArXiv Preprint ArXiv:2407.08082*. (2024), http://arxiv.org/abs/2407.08082
23. Androjna, A., Pavić, I., Gucma, L., Vidmar, P. & Perkovič, M. AIS Data Manipu- lation in the Illicit Global Oil Trade. *Journal Of Marine Science And Engineering*. **12**, 6 (2024)
24. Anuoluwapo, A. Quality Assessment of Maritime AIS Data. (2023)
25. Yang, M. Big data: issues, challenges, tools, Thesis (2019), http://www.theseus.fi/handle/10024/267608